# Sharing and Compatibility Studies between International Mobile Telecommunications Systems (Mobile Cellular) and Earth-Exploration Satellite Service (Active) in 10-10.5 GHz

Heykel Houas

ANFR, France

*E-mail: heykel.houas@anfr.fr*

July 25, 2024

## Abstract

The World Radiocommunications Conference 2023 (WRC-23), held in Dubai, United Arab Emirates, between November 2023 to December 2023, intensively discussed several Agenda Items related to possible new *International Mobile Telecommunications (IMT)[1]* identifications (under the Mobile Service allocations in the International Radio Regulations) of several radio frequency bands, utilized by other incumbent radio services. One example of a frequency band is 10-10.5 GHz, which has a primary allocation to Radiolocation, Earth-Exploration Satellite services. In order to ensure continued protection of such incumbent services (from the incoming service), sharing and compatibility studies were conducted, and debated at length, at the ITU-R Study Groups from the year 2020 to 2023.

This document presents a coexistence study, which was contributed to ITU-R Study Groups and served as a basis for the final decision making of the decision to possibly allocate 10-10.5 GHz frequency band to IMT at WRC-23. The aim of this document is to assist a follow-up journal paper which utilises technical elements described in this document. For a wider context, the coexistence study reported in this document was driven through multiple sources of inspiration:
- The need to protect radio communications systems/applications dealing with Science Services, e.g., Earth Exploration Satellite used for Synthetic Aperture Radar (SAR) operations. As the frequency range 9.2 – 10.4 GHz (a.k.a., X-band) enables improved spatial resolution (< 25 cm), improved sensitivity levels for smaller-sized objects and reduced atmospheric attenuation, it is essential for monitoring natural disaster events where cloud cover (e.g., the Eruption of Stromboli volcano in 2022) or inclement weather would otherwise hinder data collection.
- One compatibility study submitted to ITU-R Working Party 5D (5D/1466) [1].
- During the 41st Meeting of Permanent Consultative Committee II: Radiocommunications (PCC II) of the Inter-American Telecommunication Commission (CITEL) (in Region 2 of the ITU-R), one administration modified its previous Draft Inter-American Proposal (with the intention of submitting it to WRC-23), to include additional protection to the incumbent services (e.g., EESS (Active)). The same administration introduced three conditions meant to ensure protection of the incumbent services, in particular for the EESS (Active) and Radiolocation Service. Such conditions were:
   a) Administrations shall take practical measures to ensure the transmitting antennas of outdoor IMT base stations (BSs) are normally pointing below the true horizon when deployed within 10-10.5 GHz; additionally, the mechanical pointing needs to be at or below the horizon;
   b) Administrations shall use side lobe suppression techniques providing 29.5 dB of attenuation for elevation angles above 30 degrees where 0° relates to the horizon and 90° to the zenith, referenced to the maximum antenna gain at the boresight;
   c) The maximum equivalent isotropically radiated power (e.i.r.p.) emitted by an IMT base station shall not exceed 32 dB W/100 MHz;

---

[1] As indicated in the title of the document, the term "*IMT*" is the ITU-specific terminology of mobile cellular systems, which denotes a family of systems, e.g., IMT-2000, IMT-Advanced, and IMT-2020, respectively, representing the successive generations of IMT technology/standards.

It was then proposed to assess the protection of SAR satellite(s) within 10-10.4 GHz under those assumptions of IMT BS deployment i.e., considering that any antenna steers the beam below the horizon, analyzing the weighting coefficient matrix of the IMT BS antenna array (in the form of a uniform planar array) reducing its sidelobes and satisfying the 32dBW/100 MHz maximum EIRP limit.

In what follows, we describe the technical and operational characteristics of both the IMT and EESS systems, the methodology of the sharing study and the results obtained by performing the sharing study.

# 1 Technical characteristics

## 1.1 Technical and operational characteristics of IMT systems operating in the frequency band 10-10.5 GHz

The study considered outdoor suburban and urban IMT characteristics in the 10-10.5 GHz as those components were considered to be primarily influencing the potential interference levels at the SAR receivers relative to indoor IMT deployments.

### 1.1.1 Technical characteristics of IMT systems in the 10-10.5 GHz frequency band

Document 5D/716 [2] Annex 4.4 of the Chairman's Report provides the characteristics of the assumed active antenna system (AAS) antenna to be used in conjunction with the radiation pattern given in Recommendation ITU-R M.2101-0 (Annex 1 section 5) [3], when generating emissions levels of the IMT base stations. These are described below:

TABLE 1

**IMT BS antenna array characteristics in 6 425-10 500 MHz**

|   |   | Small cell outdoor/ Micro urban |
|---|---|---|
| **1** | | **Base station antenna characteristics** |
| 1.1 | Antenna pattern | Refer to Recommendation ITU-R M.2101 Annex 1, section 5 |
| 1.2 | Element gain (dBi) **(Note 1)** | 5.5 |
| 1.3 | Horizontal/vertical 3 dB beamwidth of single element (degree) | 90º for H<br>90º for V |
| 1.4 | Horizontal/vertical front-to-back ratio (dB) | 30 for both H/V |
| 1.5 | Antenna polarization | Linear ±45º |

| | | Small cell outdoor/ Micro urban |
|---|---|---|
| 1.6 | Antenna array configuration (Row × Column) **(Note 2)** | 8 × 8 elements |
| 1.7 | Horizontal/Vertical radiating element spacing | 0.5 of wavelength for H, 0.5 of wavelength for V |
| 1.8 | Array Ohmic loss (dB) **(Note 1)** | 2 |
| 1.9 | Conducted power (before Ohmic loss) per antenna element (dBm) **(Note 5)** | 16 **(Note 6)** |
| 1.10 | Base station maximum coverage angle in the horizontal plane (degrees) | ±60 |
| 1.11 | Base station vertical coverage range (degrees) **(Notes 3, 4, 6)** | 90-120 |
| 1.12 | Mechanical downtilt (degrees) **(Note 4)** | 10 |

**Note 1**: The element gain in row 1.2 includes the loss given in row 1.8. This means that this parameter in row 1.8 is not needed for the calculation of the BS composite antenna gain and e.i.r.p.

**Note 2**: 16 × 8 means there are 16 vertical and 8 horizontal radiating elements. In the sub-array case, one implementation is 2 vertical radiating elements combined in a 2 × 1 sub-array.

**Note 3**: The vertical coverage range is given in global coordinate system, i.e. 90° being at the horizon.

**Note 4**: The vertical coverage range in row 1.11 includes the mechanical downtilt given in row 1.12.

**Note 5**: In sharing studies, the transmit power calculated using row 1.9 is applied to the typical bandwidth.

**Note 6:** In sharing studies, the UEs that are below the coverage range can be considered to be served by the "lower" bound of the electrical beam, i.e. beam steered towards the max. coverage angle. A minimum BS-UE distance along the ground of 35 m should be used for urban/suburban and rural macro environments, 5 m for micro/outdoor small cell, and 2 m for indoor small cell/urban scenarios.

Notes accompanying the table generally clarify the usage of several parameters in the radiation pattern of the composite antenna calculus. However, it's worth mentioning two additional points:

- Note 3 indicates that the vertical coverage range of angles (90-120°) for suburban and urban cases relates to the *Global Coordinates System* (*GCS*) while Note 4 indicates that these values cover the mechanical tilt, which would mean that these values are not given in GCS but in Local Coordinate System of the BS antenna. Hence, it's important to retain Note 3 and disregard Note 4 when considering these values.

- Since the proposed values of the BS antenna mechanical down-tilt is positive (10°), it is assumed that any rotation in positive down-tilt[2] is performed in clockwise direction[3].

*GCS* is used to locate the SAR satellite while LCS relates to the BS antenna panel reference. Because of the mechanical down-tilting of the BS antenna and since the antenna sector is not necessarily oriented « in front » of the space station[4], geometrical transformation from one coordinate system to the other one is needed to derive the direction where to calculate the gain of the BS antenna. This operation is a combination of two rotations, the first one being the mechanical tilt and the second being the azimuth. If different BS of the same environment (urban or suburban) operate with the same mechanical down-tilt, their different azimuthal positioning with respect to the EESS active satellite will lead to different transformations to get in LCS.

---

[2] To go from LCS to GCS.

[3] In particular when deriving any gain of the BS antenna (in LCS).

[4] i.e. same azimuth of the airborne radar if another reference is taken (e.g. North) or 0° if the radar is the reference of the azimuth axis.

### 1.1.2 Operational characteristics of IMT systems in the 10-10.5 GHz frequency band

#### 1.1.2.1 BS deployment characteristics

Tables 2 and 3 below provide the deployment-related parameters of IMT systems for the frequency band 10 000-10 500 MHz. Urban and suburban micro scenarios are considered in this study.

TABLE 2

**Deployment-related parameters for bands between 10 and 11 GHz**

|  | Urban/suburban hotspot (outdoor) |
|---|---|
| **Base station characteristics/Cell structure** | |
| Deployment density (**Note 1**) | 30 BSs/km$^2$ urban / 10 BSs/km$^2$ suburban |
| Antenna height | 6 m |
| Sectorization | Single sector |
| Downtilt | See Table 10 |
| Frequency reuse | 1 |
| Indoor base station deployment | n.a. |
| Indoor base station penetration loss | n.a. |
| Below rooftop base station antenna deployment | 100% |
| Typical channel bandwidth | 100 MHz |
| Network loading factor (base station load probability X%) (see section 3.4 below and Rec. ITU-R M.2101 Annex 1, section 3.4.1 and 6) | 20% |
| TDD / FDD | TDD |
| BS TDD activity factor | 75% |

**Note 1**: These density values are for small dense areas.

#### 1.1.2.2 BS deployment in small area ($R_a/R_b$)

The following table provides the Ra/Rb factors that are considered for the study. As shown in the figure below, the area considered for the IMT deployment is divided into three zones in order to model different densities, based on Document 5-1/406 Annex 1 [4]. The total area considered is 2 000 km$^2$. In the central zones (surface smaller than 1 000 km$^2$), the Rb factor is set to 100%. For large area deployments (surface larger than 1 000 km$^2$) option 1 of Table 3 has been considered (where Rb is set to 5% for areas smaller than 200 000 km$^2$).

FIGURE 1

**Illustration of IMT deployment zones**

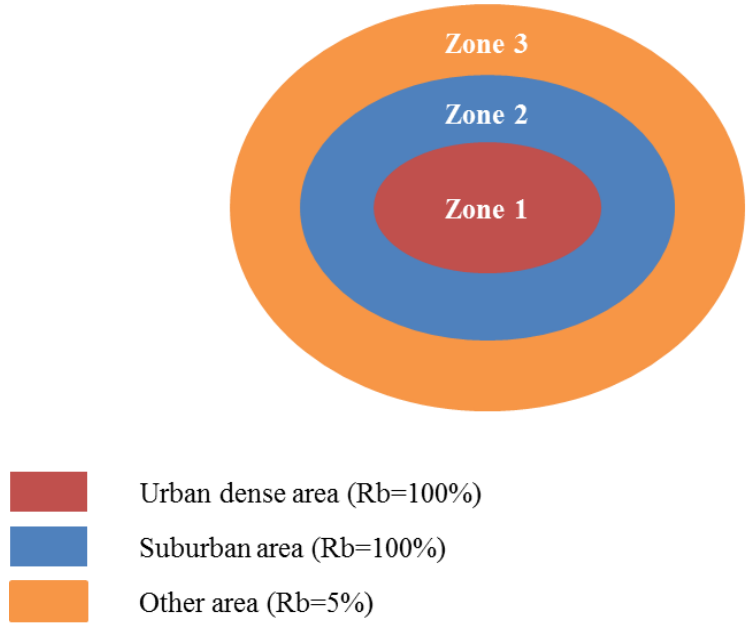

- Urban dense area (Rb=100%)
- Suburban area (Rb=100%)
- Other area (Rb=5%)

TABLE 3

**Values for Ra and Rb to be used in studies involving IMT deployments for frequency bands between 10 and 11 GHz for usage by Region 2**

|  | Option* | Hotspots (outdoor) |
|---|---|---|
| Ra | 1 | 7% Urban<br>3% Suburban |
| Rb (depending on the area under study) | 1 | 5% (area < 200 000 km$^2$)<br>2% (200 000 - 1 000 000 km$^2$)<br>1% (area > 1 000 000 km$^2$) |
|  | 2 | 2.5% (area < 200 000 km$^2$)<br>2% (200 000 - 1 000 000 km$^2$)<br>1% (area > 1 000 000 km$^2$) |

## 1.2 Technical and operational characteristics of EESS (active) operating in the frequency band 10-10.5 GHz

### 1.2.1 Technical characteristics

Recommendation ITU-R RS.2105-0 [5] provides the "*Typical technical and operational characteristics of Earth exploration-satellite service (active) systems using allocations between 432 MHz and 238 GHz*". One particular sensor system operates in 10-10.4 GHz. The characteristics of this sensor are demonstrated in the table below for a *Sun Synchronous Orbit (SSO)*.

TABLE 4

**Characteristics of EESS (active) Sensor SAR-F6 in 10 – 10.4 GHz**

| Parameter | SAR-F6 |
|---|---|
| Sensor type | SAR |
| Type of orbit | Circular, SSO |

| Parameter | SAR-F6 |
|---|---|
| Altitude (km) | Varying |
| Antenna type | Active phased array |
| Number of beams | 1 |
| Antenna (Receive) peak gain (dBi) | 47 |
| Antenna beam look angle (degrees) | 18-50 |
| Antenna beam azimuth angle (degrees) | 90 |
| Antenna elevation beamwidth (degrees) | 1.13 |
| Antenna azimuth beamwidth (degrees) | 0.53 |
| RF centre frequency (MHz) | 9 800 |
| RF bandwidth (MHz) | 1 200 |
| System noise figure (dB) | 3 |
| Total Integrated Gain TIG (dB) (**Note 1**) | 0.25 |
| Efficiency of the antenna (%) (**Note 1**) | 70 |

**Note 1**: The TIG of active antenna should be in reality below 0 dBi due to the effect of efficiency, but the only way to take that element into account is to apply the correction factor of $10 \cdot \log_{10}(\text{efficiency}) - 0.25$ dB to the entire pattern. For area in the main beam of the SAR satellite such correction factor should not be used. Therefore, in the study the TIG and the efficiency of the antenna is taken into account only in zone 2 and 3.

It's worth noticing that the same recommendation states that *EESS (active) systems operate in non-geostationary satellite orbit (non-GSO). Orbits are typically circular with an altitude between 350 and 1 400 km. Some EESS (active) systems operate in a sun-synchronous orbit. Some sensors make measurements over the same area on the Earth every day, while others will repeat observations only after a longer (often more than two weeks) repeat period.*

### 1.2.2  Protection criterion of EESS active

Recommendation ITU-R RS.1166-4 [6] provides "*Performance and interference criteria for active spaceborne sensors*" in the EESS (active) allocations between 432 MHz and 238 GHz.

Table 2 under *recommends* 2 of Recommendation ITU-R RS.1166-4 provides "the interference and data availability criteria given in Table 2 be applied for instruments used for active sensing of the Earth's land, oceans and atmosphere". For SAR systems, it is as follows:

TABLE 5

**Protection criteria for EESS (active) service in 10 – 10.4 GHz band**

| Sensor type | Interference criteria | | Data availability criteria (%) | |
|---|---|---|---|---|
| | **Performance degradation** | *I/N* (dB) | **Systematic** | **Random** |
| Synthetic aperture radar | 10% degradation of standard deviation of pixel power | –6 | 99 | 95 |

As shown in the table above, the SAR protection criteria is given under the form of *I/N*=-6dB not to be exceeded for more than 1%.

## 1.2.3 Receiving SAR satellite antenna pattern

As presented in Table 4 above, the antenna *Beam Looking Angles (BLA)*, equivalent to off-nadir angle of the SAR satellite is between 18 and 50 degrees.

The SAR F6 uses a spotlight mode of operation as defined in Recommendation ITU-R RS.2043. The most common pointing practice is having the satellite spotlight pointing orthogonally to the satellite track. In other words, in the case of a highly inclined orbit with a ground track from south to north, the SAR satellite would point eastward or westward. The following figure provides an illustration of such a mode of operation:

FIGURE 2

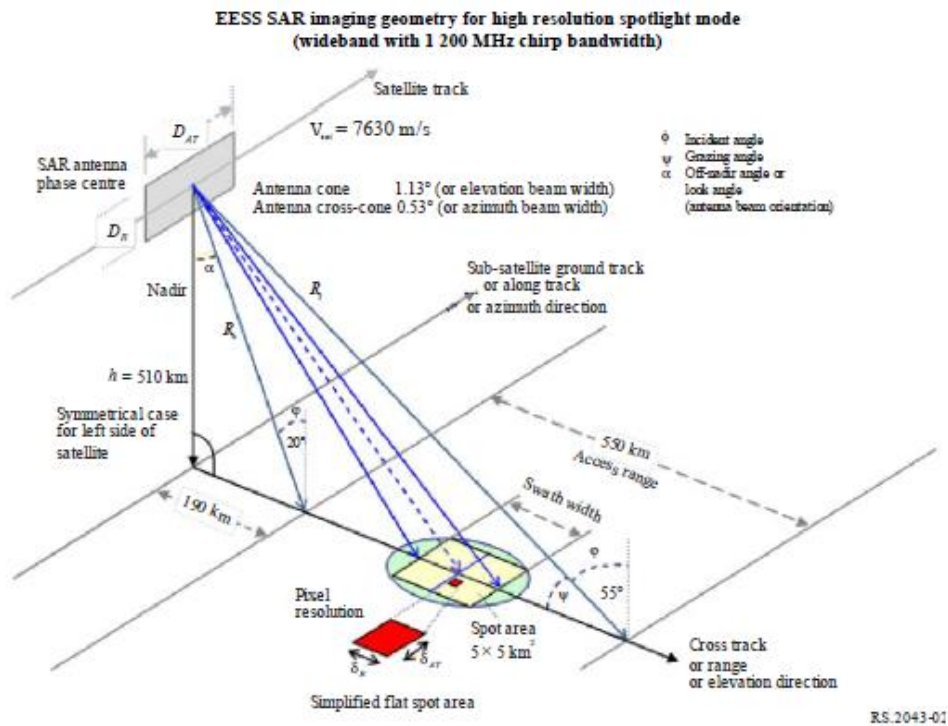

The Table 9 of Recommendation ITU-R RS.2043 provides the average SAR gain as a function of azimuth and elevation. In addition, the liaison statement Document 5D/573 [7] from WP 7C to WP 5D clarifies the definition of the "horizontal" and "vertical" angles to be used in Table 9 of Recommendation ITU-R RS.2043. The following figure provides an illustration of the receiving SAR satellite antenna pattern used in the study presented below.

FIGURE 3

**SAR antenna gain over whole sphere**

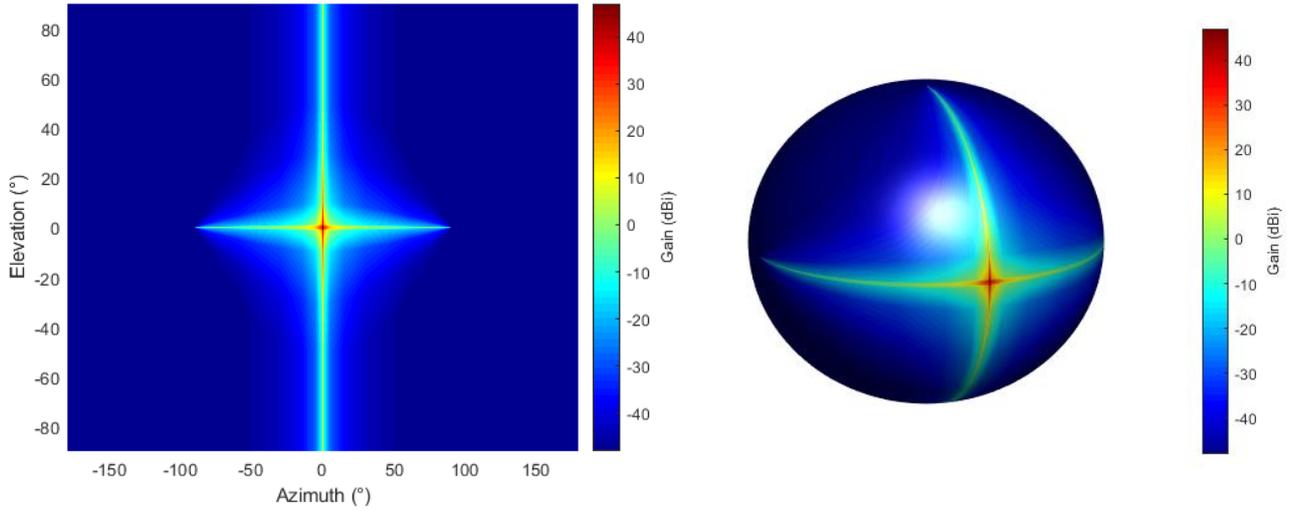

The following figure provides the SAR satellite footprint for different BLAs (i.e.m off-nadir angle) of the space station antenna: 18° (left figure) and 50° (right figure) while pointing towards the East in azimuth. For the sake of readability, the antenna gain of the space station ranges down to -10dBi. In order to achieve a footprint centred in Sao Paulo city's downtown (i.e., around (-46.3°E, -23.3°N)) for different BLAs, different (longitude, latitude) positions within the orbit of the space station have been considered: (-48.0187°E, -23.6060°N) for BLA=18° and (-52.8908°E, -23.4922°N) for BLA=50°.

FIGURE 4

**SAR satellite antenna footprint with beam look angles of 18° (left) and 50° (right)**

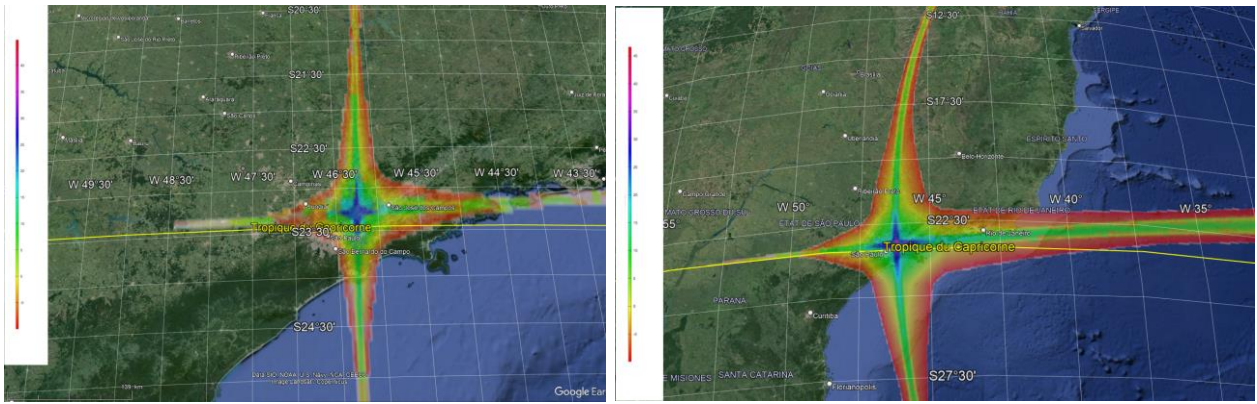

## 1.3 Propagation assumptions between IMT and SAR satellite systems

The propagation losses are based on Recommendation ITU-R P.525 [8] for the free space loss component of the slant path separating a space station from a ground station.

The attenuation due to the clutter environment around the IMT hotpots (located at B in the below picture) have also been considered using the Earth-to-space clutter model described in Recommendation ITU-R P.2108 [9] (see distribution in the following figure). Random percentages of locations (0-100%) were considered for this analysis to depict the variability of such losses in connection to the proximity and the size of the clutter surrounding the transmitting BSs. Boresight elevation angle α from the ground station to the satellite (depicted in S in the below picture) has been calculated from BLA parameter ($\beta$ angle in the below figure) using spherical Earth model (α=34.43° for BLA=50° and α=70.57° for BLA=18°), noting that an Ellipsoid model would have given a very close value (e.g. for BLA=18°, α = 70.569° with ellipsoid model and 70.567° with spherical model).

FIGURE 6: **Relationship between BLA β and elevation angle α for the spherical Earth model**

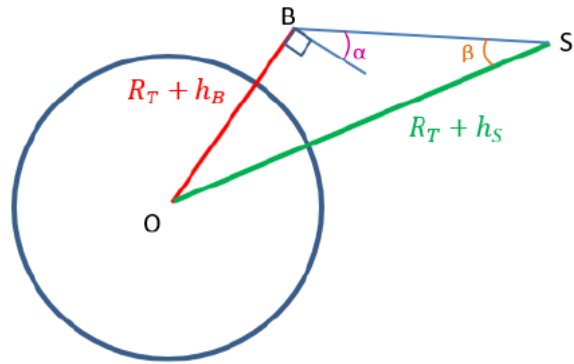

FIGURE 7

**Probability Distribution Function and Cumulative Distribution Function of the resulting clutter loss at beam look angle of 50° (i.e. α=34.43°)**

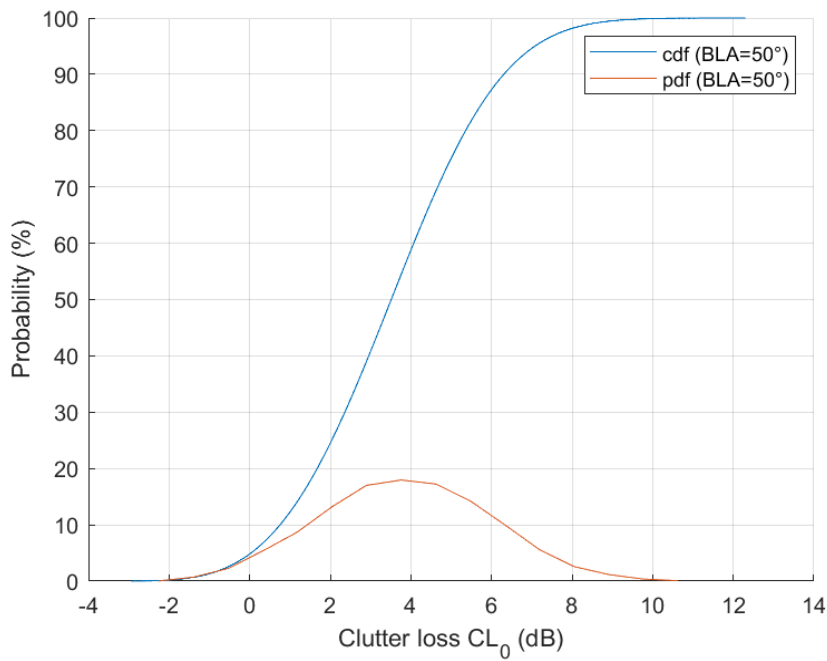

## 2 Methodology for the sharing analysis

### 2.1 Static analysis of the interference received by SAR satellite

As a non-geostationary satellite, a SAR space station can cover a number of missions in relation of the mapping of a given area by using an antenna phase array enabling beamforming. The planar array can be controlled in order to steer the beam in any direction either in elevation or/and in azimuth depending on the nature mission as mentioned in ITU-R Report RS.2314-0[5] [10]. The Scan Mode is dedicated for the coverage of large area e.g. 840 km x 100 km like maritime use cases where national security authorities might need persistent visibility (e.g. to identify security threats or illegal activities occurring in country borders). For this scenario, azimuth scanning remains constant while simultaneous BLAs are used to provide a wide coverage through multiple strips. The Strip mode is suitable for detecting changes to vast sea and land areas like natural disasters (flood detection, volcano eruption) or illegal activities (very thin oil films floating on the sea surface) the azimuth of the beam is left unchanged and a single BLA is performed. Finally, the spotlight mode focus on detailed monitoring of a smaller area e.g. 15 km x 15km by creating multiple images of the same area (called scene) for a time evolution of its pattern. In order to track the region of interest, SAR antenna must be steered in azimuth and in elevation in order to track the region of interest.

Therefore, the trajectory of the satellite is needed in order to evaluate at which position it is able to perform a mapping of the region of interest. Recalling that its orbit can be featured through multiple parameters such as the shape and size of the ellipse: eccentricity $e$, altitude at the periapsis[6] $h$, the orientation of the orbital plane in which the trajectory is contained: the inclination $i$, longitude of the ascending node called the *Right Ascension of the Ascending Node* (*RAAN*) and finally the position of ellipse within its orbital plane/of the space station within its trajectory: argument of periapsis and true anomaly. Time aspect in Non-Geostationary satellite is also important to investigate the suitable position of the satellite when several mappings of the same area under different beam-steering are built: sampling level (1s per position) as well as duration of the operation as the SAR can pass over a contiguous territory multiple times per day. Below table and figure capture those driving assumptions as output of discussion with a SAR satellite manufacturer. Figure 8 captures the varying nature of a non-geostationary satellite altitude above the sea level.

TABLE 6

**SAR satellite orbit parameters**

| Name | Value |
|---|---|
| Altitude at the periapsis (km) | 240.2753 |
| Inclination $i$ (degrees) | 88.0008 |
| RAAN $\Omega$ (degrees) | 273.9587 |
| Eccentricity e | 0.0202 |
| True anomaly $\nu$ (degrees) | 180.4155 |
| Argument of perigee $\omega$ (radians) | 180.9383 |

FIGURE 8

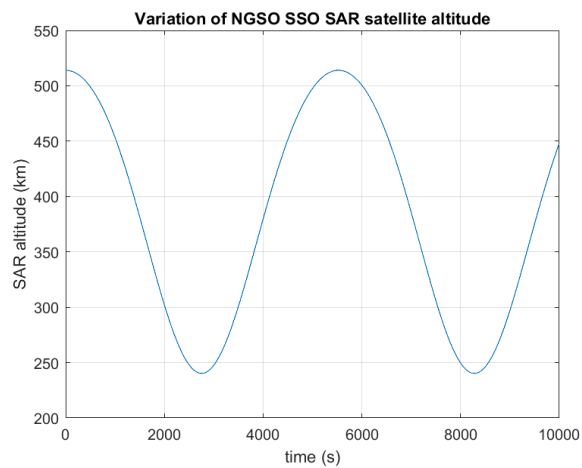

From these parameters, the orbit of a SAR satellite can be derived and is plotted as follows (red curve) in the below figure. Indeed, in case a specific area is to be mapped, it is possible for the space station to perform an imaging by adjusting its BLA depending on its position with respect to the area of interest. Recalling that the beam pointing of antenna is directed to the East, one could notice in light of SAR satellite western positions with respect to San Paulo (yellow point) that higher BLA (50°) is required for the SAR satellite location farther (green) than Sao Paulo while the closest position (black) fits for BLA=18°.

---

[5] See Figure 2.

[6] Corresponding to the lowest altitude of the satellite within its orbit.

FIGURE 9: **Orbit for EESS Active satellite**

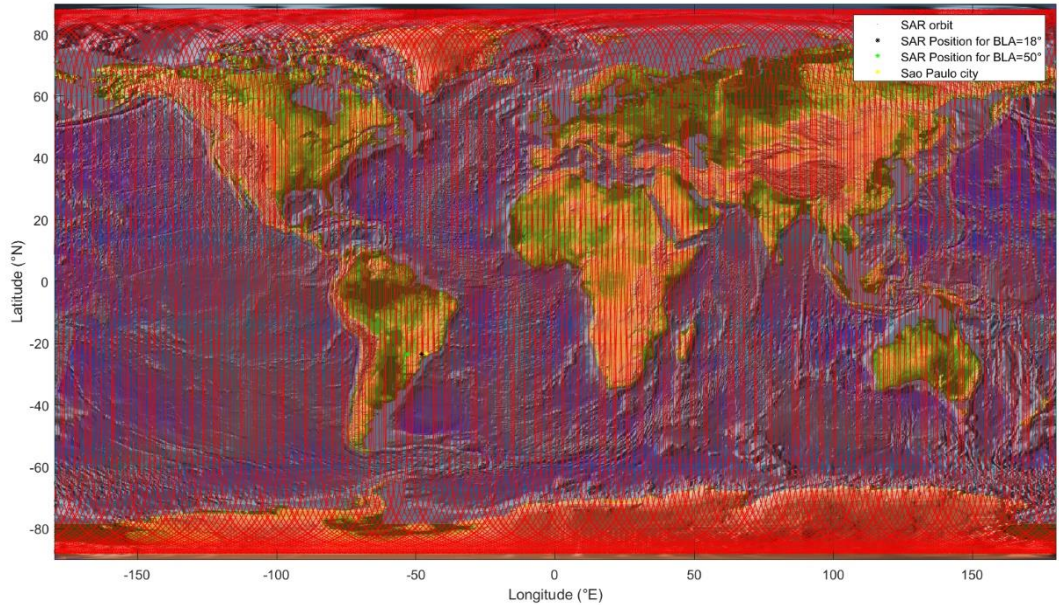

This analysis considers a fixed position of the SAR satellite (within its orbit) in view of a given area over the Earth where it performs its mapping of Sao Paulo city (around 1500 km$^2$) and its vicinity (around 500 km$^2$). Within the given footprint where the space station operates at a given BLA, IMT BS hotspots are deployed within the in-land area and would behave as interferers when actively transmitting. As highlighted previously, there is a statistical dimension into the aggregate interference to be received by the satellite receiver at a fixed orbit:
- The beam-steering of every active BS (to serve a given UE dropped within the cell covered by the BS) within the served cell;
- The attenuation due to clutter environment as a function of the location variability of the BS location with respect to surrounding buildings.

Noting that BSs within the footprint of the SAR satellite main beam remain the same if the geographical position of the space station is fixed, one could conclude that this study is a static one where only statistical aspects of the parameters driving the interference were tackled. Even though the set of beam-steering raw values (of any BS) is a time sampling of the beam orientation, one could also reorganize this vector of observations as a statistical distribution (without specific order) if the (beam-steering) behaviour of every BS antenna is similar. In such case, stochastic process related to BS antenna gain $G_{Tx}(\omega, t)$ (where $\omega$ relates to the outcome, $t$ refers to the time instant) can be considered as ergodic if any calculation involving $G_{Tx}$ (mean, auto-covariance, etc.) over possible outcomes $\omega \in \Omega$ is equal to the calculation involving $G_{Tx}$ (regarding the same moment) over time $t$.

Consequently, aggregate interference assessment for a receiving space station at a fixed location can be described a **static** analysis.

**2.2       Approach to compute the aggregate interference received by SAR satellite**

This section presents a methodology to estimate the aggregate interference into an EESS active SAR satellite at a fixed location by IMT-2020 BS deployment within the main beam of that SAR satellite antenna (3dB beamwidth) when pointing towards a dense area such big cities.

The goal is to assess the impact of all active (i.e. simultaneously transmitting) IMT BS into a SAR satellite antenna main beam.

**Step 1:** Generating random IMT base stations deployment based on a uniform distribution in surface (as this relates to hotspots that are not structured in grid like Macro BSs) in 3 contiguous zones around the SAR satellite main beam area (based on Figure 1) as defined below:(surface larger than 1 000 km2) option 1 of Table 3 has been considered (where Rb is set to 5% for areas smaller than 200 000 km2).Z1: Zone 1 is the central zone and it is considered to be a dense urban area, with a size of 200 km$^2$.

Z2: Zone 2 is the ring around zone 1 and it is considered to be a dense suburban area. The total area zone 1 + zone 2 is set to 1 000 km$^2$, and therefore the size of zone 2 is 800 km$^2$.

Z3: Zone 3 depends on the total surface of the area of deployment. In this study, the total deployment area, i.e. Zone 1 + Zone 2 + Zone 3 is set to 2 000 km$^2$. The size of zone 3 is therefore equal to 1 000 km$^2$.

The number of IMT base stations that has been used in the study in each zone, is calculated based on Document 5-1/406 Annex 1 [4]. The method is recalled as follows:

− Considering a coverage area for the SAR satellite at the lowest elevation (34.43° derived from the maximum BLA of 50°) for worst-case analysis. The results are also computed for BLA=18 deg.

− There is a maximum density of 30 BSs/km$^2$ in the urban environment and 10 BSs/km$^2$ in the suburban environment. With this BS density information combined with the Ra/Rb methodology, the BS network load factor and TDD factor, the total number of BS simultaneously transmitting within the 3 different zones can be calculated using the following formulas:

$$N_{BS\_Z1} = S_1 \times BS_{AF} \times BS_{NLF} \times R_{aU} \times D_{BS_U}$$

$$N_{BS\_Z2} = S_2 \times BS_{AF} \times BS_{NLF} \times R_{aSU} \times D_{BS_{SU}}$$

$$N_{BS\_Z3} = S_3 \times BS_{AF} \times BS_{NLF} \times 0.05 \times (R_{aSU} \times D_{BS_{SU}} + R_{aU} \times D_{BS_U})$$

with

$BS_{AF}$ BS TDD activity factor (75%);

$BS_{NLF}$ BS network loading factor (20%)

$S$ Z1, Z2 or Z3 surface (km$^2$);

$Ra_{SU}$ the ratio of suburban hotspot coverage areas to areas of cities/built areas/districts (=3%);

$Ra_U$ the ratio of urban hotspot coverage areas to areas of cities/built areas/districts (=7%);

$Rb$ for zones 1 and 2 the ratio between built areas to total area is taken to 100%, for zone 3 this ration is taken to 5% (see option 1 in Table 2);

$D_{BS_{SU}}$ BS density in the outdoor suburban hotspot (10 BS/km$^2$);

$D_{BS_U}$ BS density in the outdoor urban hotspot (30 BS/km$^2$).

In this study, we are considering the overlap between IMT and EESS (active) in the band 10-10.4 GHz.

The nominal results have been performed for the case in which there are four operators, each one using 100 MHz, occupying together the 400 MHz between 10 – 10.4 GHz in co-channel with EESS (active) sensors. To simulate this effect, the number of base stations $N_{Base\ Stations}$ has been multiplied by 4 and the noise bandwidth has been increased to 400 MHz.

– For the lowest SAR satellite elevation pointing case, for $Z1=200$ km$^2$ and network loading factor of 20%, a total of 63 urban BS are transmitting simultaneously co-frequency in Z1, of 36 urban BS in Z2 and 4 urban and suburban BS in Z3.

**Step 2:** Calculating the BS gains towards the satellite victim receiver.

Each BS pointing direction is based on the following UE deployment characteristics:

– Electrical azimuth pointing (phi scan): normal distribution Mean 0° standard deviation 30° and truncated at +/- 60° (see distribution in Figure 8);
– Distance: Rayleigh distribution with a standard deviation of 32 m (see distribution in Figure 10);
– Elevation: computed from the azimuth and distance distributions with a BS at 6 m altitude.

FIGURE 10

**Rayleigh distribution used for the generation of UE in distance**

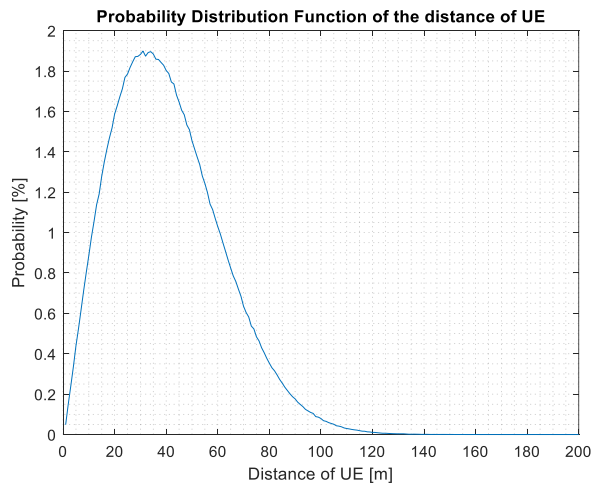

It is important to note that the BS vertical coverage range is limited to 90-120° for both the suburban and urban deployment cases, which is equivalent to limiting the elevation angle from 0° (horizon) to -30° (below the horizon). Following Table 2, for all UE deployments seen at elevation angles lower than -30°, the elevation angle of the BS to cover these UEs will be set to -30° and the UE will be covered with the BS side-lobes.

The elevation is a combination of both electrical tilt and mechanical tilt. The electrical tilt can be determined by removing the mechanical down-tilt contribution (10°) from the obtained elevation pointing distribution.

FIGURE 11: **distribution of AAS beam-steering angles when serving UEs within the small-cells**

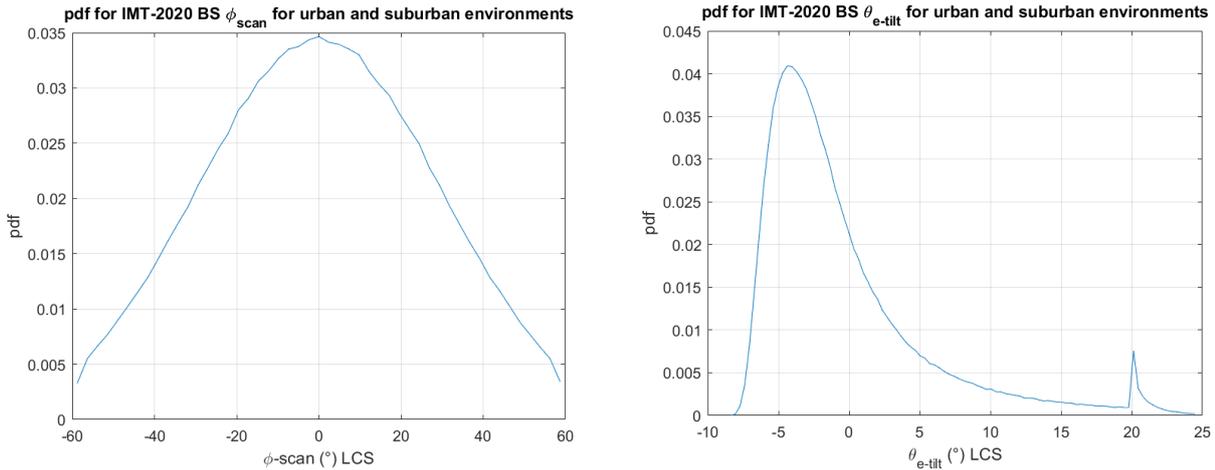

Based on the parameters set by WP 5D and using formula given in Rec ITU-R M.2101-0, the distribution of BS antenna gain for different elevation angles towards the SAR satellite (*i.e.* $\varphi = 0°$ in GCS) is generated over 50 000 runs (i.e 50 000 random directions of the beam are built over the cone of coverage – delimited by horizontal -60..60° and vertical bounds 90..120° in GCS - ) displayed for suburban and urban environments in terms of *complementary cumulative density function (c-cdf)*: It should be noted that the location of azimuth BSs antenna panel with respect to the space station is (uniformly) randomly generated within -180..180 range. This means that the BS antenna gain is calculated <u>over the whole BS azimuth range, i.e. -180 to 180 degrees</u>.

FIGURE 12: **Distribution of BS antenna gain towards EESS satellite**

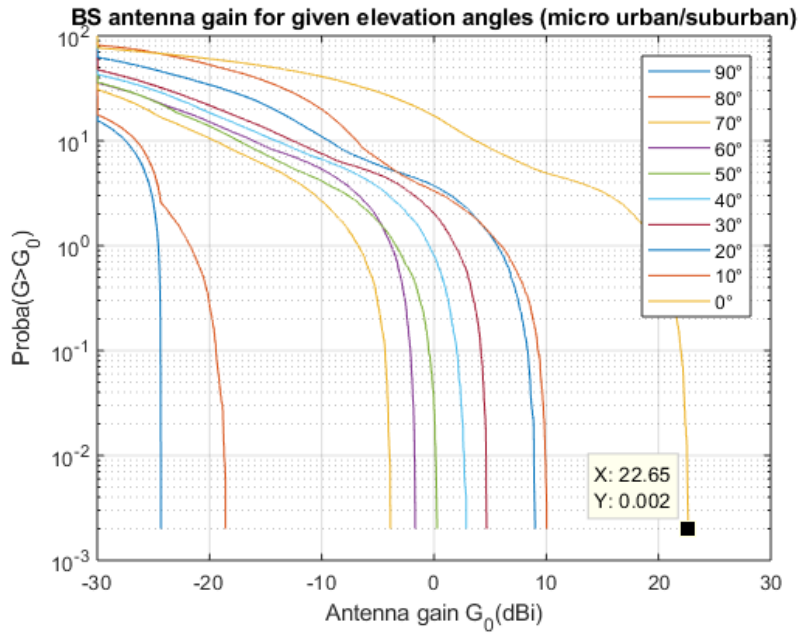

One could find out:
- That for higher elevation angles (0° related to the horizon, 90° related to the zenith), the quantile for which there is a percentage exceedance is decreasing. Such fact comes from the down-tilt effect of the micro BS which results in lower gain above the horizon.

- The lowest value of the curve (related to $2.10^{-2}$ percent) matches with the number of samples[7] (50 000).

Finally, the radiated power of micro BS (in urban and suburban areas) can be calculated by adding the conducted power (from parameters of Table 1) using to the BS antenna gain derived earlier. It should be noted that due to multiple UEs (3 per sector) served by the same micro BS sector, the conducted power of micro BS is equally apportioned (as channel bandwidth is equally shared by the three terminals) when radiating the power to three different terminals. Accounting 2dB extra power proposed by several administrations during WRC-23, Peak EIRP equates: $16 dBm + 2 dB + 10 log_{10}(8 \times 8 \ single \ elts \times 2 \ polarizations) + 5.5 + 10 log_{10}(8 \times 8) = 62.6 dBm$ while peak value of the statistical EIRP per user is derived by using the peak value of the BS antenna gain distribution (c-cdf) and apportioning the conducting power among those three UEs:

$$16 dBm + +2 dB + 10 log_{10}\left(\frac{8 \times 8 \ single \ elts \times 2 \ polarizations}{3 \ users}\right) + 22.65 dBi \approx 57 dBm$$

Total Radiated Power (TRP) can be calculated using the above information as follows:

$$16 dBm + 2 dB + 10 log_{10}(8 \times 8 \ single \ elts \times 2 \ polarizations) \approx 39 dBm \ (dual \ polarization)$$

$$16 dBm + 2 dB + 10 log_{10}(8 \times 8 \ single \ elts) \approx 36 dBm \ (single \ polarization)$$

Work undertaken prior WRC-19 (Document 5-1/124) [11] and also from external organizations (3GPP) indicate the need to ensure that the total array directivity is equal to 0 dB minus any loss related to the antenna array (efficiency, ohmic, coupling, mismatch and/or insertion loss…). The calculation of the *Total Integrated Gain (TIG)* over the whole sphere for the single radiating element leads to around -2 dB, indicating that the 2 dB array ohmic loss provided in above table is already included in the analytical expression of the single element radiation pattern (from the directivity of the element $D_e$).

Performing the same kind of calculation for the composite antenna where the beam-steering is generated within the vertical and horizontal coverage following the guidance from ITU-R Rec M.2101 and parameter values from Table 1 (of this document) leads to an overall range of variation -2..0 dB for *TIG* as depicted in the below picture for the suburban and urban micro environments (left side figure). This small variation of the TIG in negative values is consistent with array (mismatch, coupling, ohmic) losses affecting the composite antenna in the beamforming process. Since *Sidelobes Suppressions Levels (SSL)* methodology to be applied to BSs AAS is also considered for this study, TIG for AAS with SSL using Taylor 1 weighting coefficient (described in Doc 5D/796) is also displayed in the below picture (right side). It shows a larger range of variation for the negative values -6..-3 dB because of the non-uniform windowing[8] usage (Taylor) although the single element pattern remains the same.

---

[7] $2.10^{-2}$=1/50000

[8] Distribution of excitation signals level to the array.

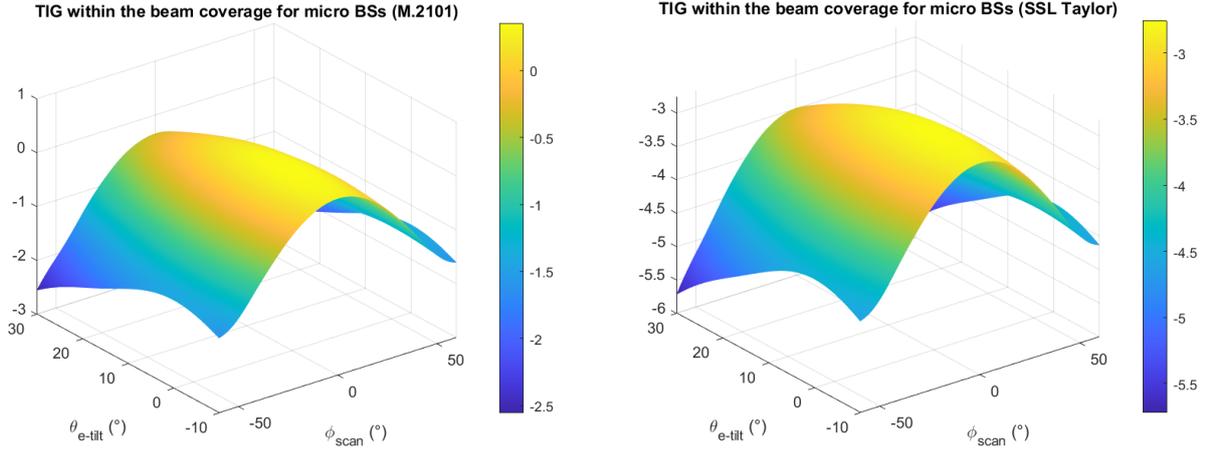

FIGURE 13: **TIG for composite AAS using different radiation pattern models**

Therefore if normalization of the composite antenna gain for the IMT micro BS is not needed for the current analytical expression from Recommendation ITU-R M.2101-0 because it captures the losses related to antenna, this is not the case for the SSL model and **a normalization factor has to be applied** as a function of the beam-steering angles couple ($\varphi_{scan}, \theta_{e-tilt}$) for every active micro BS. Following formula captures the normalization process applied to the directivity of array $D_a$:

$$D_a(\theta, \varphi, \theta_{e-tilt}, \varphi_{scan}) = 4\pi \frac{D_e(\theta, \varphi).\left|\sum_{m=1}^{N_H}\sum_{n=1}^{N_V} I_{mn}.v_{n,m}(\theta, \varphi).w'_{n,m}(\theta, \varphi, \theta_{e-tilt}, \varphi_{scan})\right|^2}{\int_{-\frac{\pi}{2}}^{\frac{\pi}{2}}\int_{0}^{\pi} D_e(\theta', \varphi').\left|\sum_{m=1}^{N_H}\sum_{n=1}^{N_V} I_{mn}.v_{n,m}(\theta', \varphi').w'_{n,m}(\theta', \varphi', \theta_{e-tilt}, \varphi_{scan})\right|^2 d\theta' d\varphi'} \quad (1)$$

Where $(I_{mn})_{1 \leq m,n \leq 8}$ refers to the Taylor matrix of weighting coefficients feeding the planar array,

$w'_{n,m}$ refers to the beam scanning component for $(m,n)^{th}$ single element of the planar array,

$v_{n,m}$ refers to the phase shift for $(m,n)^{th}$ single element of the planar array.

**Step 3:** Determining the aggregate interference form the IMT-2020 base station deployment into the SAR satellite.

− The off-axis gain for each of the IMT base station towards the SAR satellite is calculated following step 2;

− The aggregate interference is calculated using the following formula:

$$I/N_{aggregate_{dB}} = 10 \times log_{10}\left(\sum_{i=1}^{n} 10^{\frac{I/N_i}{10}}\right)$$

where $I/N_i$ is the interference over noise of the BS interferer $i$ out of the total number of BS deployed at each iteration:

$$I/N_i = P_{tx} + G_{tx_i}(\theta_i, \phi_i) + G_{rx} - PL_i - CL_i - N_{rx} - L_P$$

$P_{tx}$ is the IMT BS power spectral density for both suburban and urban BS:

$$P_{tx} = P_{el} + 2\text{dB} + 10log10(N) - 10log10(B_{width}) - 30$$

$$P_{tx} = -10{,}93 \text{ dB(W/MHz)};$$

$B_{width}$: IMT BS bandwidth (100 MHz)

$P_{el}$: the power of a single element (16 dBm)

NV: number of elements of the phased array vertically (8);

NH: number of elements of the phased array horizontally (8)

N total number of elements (equal to NV x NH x 2)

$G_{tx_i}(\theta_i, \phi_i)$ is the i-th IMT BS gain towards the SAR satellite with a beam arriving at the minimum elevation of 34.4 degrees[9];

$G_{rx}(\alpha)$ is the SAR satellite main beam gain towards each IMT BS calculated using the antenna pattern model in Recommendation ITU-R RS.2043 [12]; $\alpha$ is the off-axis angle from each BS with respect to the direction of the beam centre.

$PL$ is the propagation loss based on Rec. ITU-R P.525.

$CL$ The clutter losses based on the Earth-to-space clutter model described in Recommendation ITU-R P.2108.

$\theta_i$ and $\phi_i$ the elevation and azimuth towards the SAR satellite as seen from the i-th IMT BS.

$N_{rx}$ the SAR satellite noise level ($10 \times \log_{10}(1.38 \times 10^{-23} \times 290 \times B)$ + NoiseFigure). With B=400 MHz for baseline (4 operators).

$L_P$ Polarization loss (3 dB)

− The $I_{agg}/N$ result is stored.

**Step 4:** redo steps 1 to 3 sufficiently to obtain a stable $I/N_{agg}$ cumulative distribution function curve and store it. In this case, a total of 10 000 iterations was performed.

## 2.3   Sensitivity study with IMT BS radiation pattern with lower sidelobes

The study presented in the previous section is based on the baseline parameters and uses the IMT BS antenna pattern given in Recommendation ITU-R M.2101, which does not include any model with lower sidelobe (using a non-uniform tapering sequence feeding the single elements of the planar array). In the baseline scenario, it is considered that 4 operators are deployed in the band, each using a bandwidth of 100 MHz. This sums up to a total of 400 MHz of spectrum. If the whole 10-10.4 GHz band was not fully licensed in some countries e.g. a single mobile operator would only use 100 MHz, the result of this coexistence study would differ as the averaging process of the aggregate interference due to multiple blocks of licensed spectrum would not occur.

It is then proposed, similarly to what was carried out by several studies submitted during the cycle of study, to conduct sensitivity analysis with different scenarios:
- Case 1: there is only one operator using 100 MHz of spectrum, deployed in the scenario. All other parameters were not changed with respect to the baseline scenario (no SSL is applied to AAS).
- Case 2: there are four operators using each 100 MHz, but with AAS performing SSL to enable lower sidelobes, all other parameters were not changed to evaluate the quantitative effect of sidelobe suppression.
- Case 3: there is only one operator using 100 MHz of spectrum, deployed in the scenario but with AAS performing SSL to enable lower sidelobes.

For case 1, the aggregate interference is modelled as received over 100 MHz SAR satellite bandwidth (where noise power *N* is computed).

---

[9] Corresponding to BLA=50°.

For cases 2 and 3, as reminded in the previous section (see Step 2), the composite gain of the IMT BS antenna for each pair ($\theta_{etilt}, \varphi_{escan}$) is adjusted using the normalization factor shown in (1).

The difference between sensitivity analysis conducted during the study cycle of WRC-23 and the one from this document is that the combination of assumptions deviating from the baseline was not carried out, namely Case 3 covering the SSL implementation + occupancy of the spectrum by a single operator (100 MHz).

Although these scenarios are investigated for this analysis, it should be noted that the radiation pattern model described in Document 5D/796 [13] (enabling lower sidelobe level using Taylor windowing matrix) is a theoretical model. In contrast to mathematical models, amplitude and phase settings of real beamformer are not ideal and amplitude and phase settings are not fully decoupled, which can affect the Sidelobe Suppression Level performance from mathematical models. To understand the real impact on other systems, it would be necessary to perform tests on real antennas to ensure that 30 dB reduction from the peak gain is feasible, in particular on dynamic environments with different beam forming directions. In case of a moving beam, it can be expected that this will degrade the sidelobe levels due to the coupling of amplitude and phase of the planar array.

# 3       Study results

## 3.1      Baseline scenario

The results of the study above provide an estimation of the impact of IMT micro cell BSs deployment in America into SAR satellite receivers. The results are provided for the edge value of the SAR satellite pointing angle (i.e. beam look angle of 18° and 50°) in order to depict elevations where extrema values of aggregate interference will be received. The results in the next figure are presented for the baseline scenario for deployment. Note that the results of all scenarios have been computed using 163 840 snapshots.

FIGURE 14 **Aggregate interference for the baseline scenario**

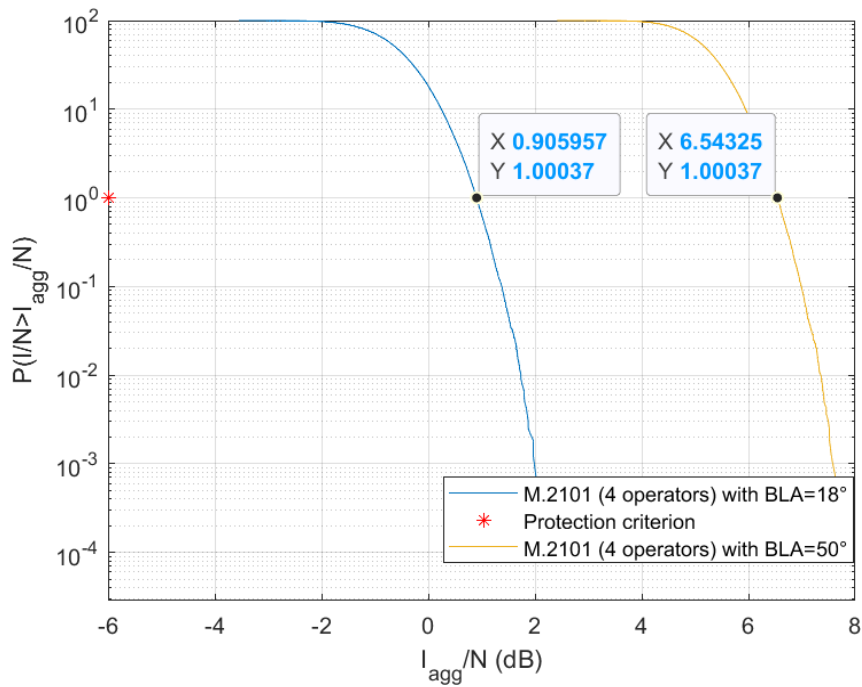

Above results show that for both BLA cases, the protection criterion (I/N=-6dB) is significantly exceeded with at least +6.9dB above this threshold. It's not surprising that the highest value is achieved for BLA=50° if it is reminded that the SAR satellite 3dB (main beam) antenna footprint increases with the BLA due to the major-axis dependence with respect to the BLA with up-to 12.5 dB (for BLA=50°).

## 3.2 Sensitivity analysis scenario

The following figure shows the results for all the sensitivity analysis scenarios and the baseline scenario for BLA = 50 deg. Note that the results of all scenarios have been computed using 163 840 snapshots.

FIGURE 15: **Aggregate interference for sensitivity analysis**

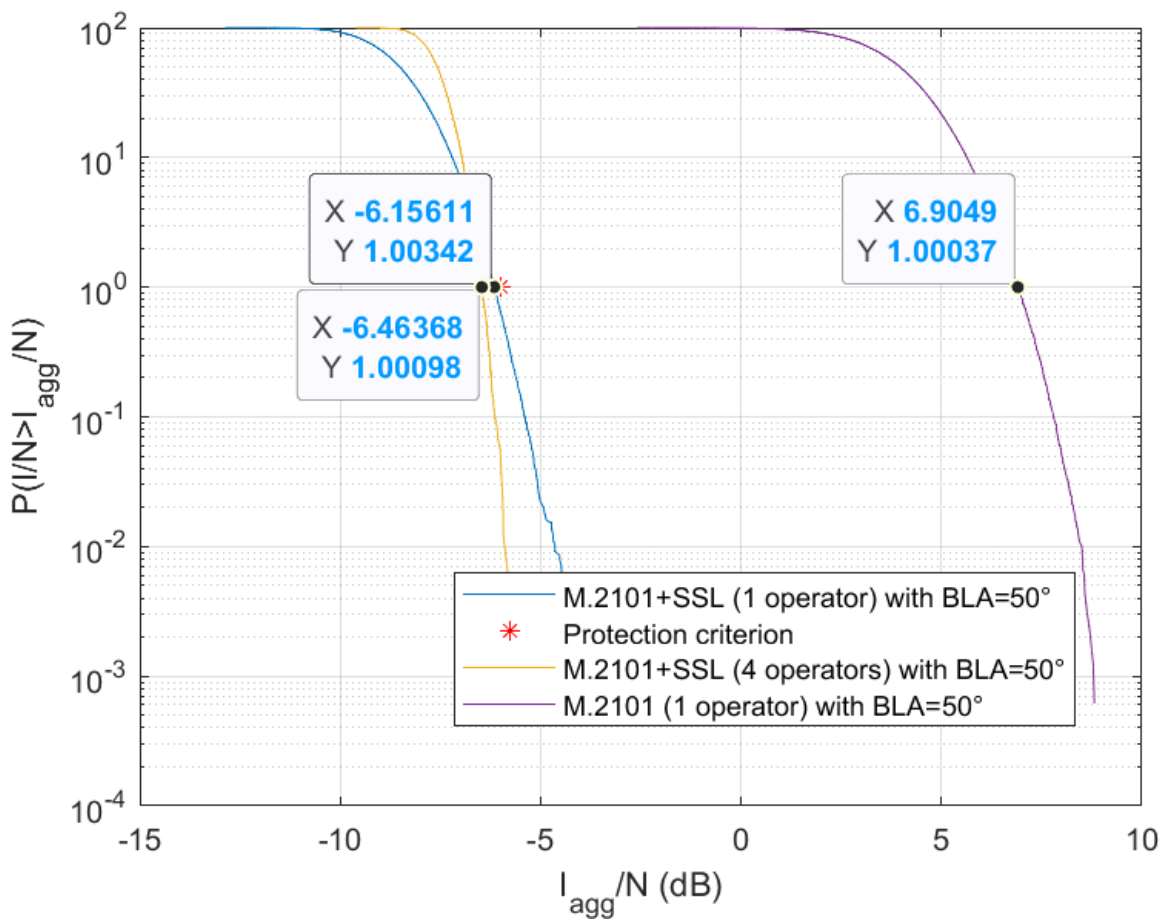

The above figure enables to draw several observations:

- The use of only 100 MHz by only one operator (Case 1) would slightly increase the aggregate interference SAR satellite would receive (0.3-0.4dB) as the averaging effect of the aggregate interference from multiple operators is lost here.
- The Sidelobe Suppression Level (Cases 2 and 3) applied to IMT BS planar array would significantly reduce the interference towards the EESS (active) systems, below the protection criterion. It should be noted that for the worst-case scenario (Case 3), there is almost no margin (around 0.1dB), suggesting that the additional 2dB increase of the peak EIRP proposed by several administrations during WRC-23 were done with the highest BS power relaxation below the horizon for coverage purpose.

# 4 Conclusions

The baseline scenario shows that the aggregate interference from BS micro cell exceeds the SAR protection criteria by 12.5 dB. These results show that sharing between IMT-2020 and SAR sensors operated in the EESS (active) in the frequency range 10-10.4 GHz (co-channel) is not feasible. Strong reduction of power of IMT BSs are required in order to protect for the long-term the operation of SAR satellites within 10-10.4 GHz, especially from 34.4° above the horizon where the active sensor would see the interference. The implementation of AAS with lower sidelobes (a minimum of 30dB lower than the peak gain starting from the first sidelobe) within IMT hotspots would theoretically protect the EESS (active) in the frequency band 10.0-10.4 GHz. However, there is no evidence that side-lobe suppression mitigation technique implementation can hold such performance while covering IMT capacity operations for the end-users, especially when serving simultaneously multiple users in the same Orthogonal Frequency-Division Multiplexing (OFDM) time-frequency resource, while maximising the Signal-to-Interference-plus-Noise Ratio (SINR).

_____________